\documentclass[a4paper]{jpconf}
\usepackage{graphicx}
\begin{document}
\title{Family of $N$-dimensional superintegrable systems and quadratic algebra structures}

\author{Md Fazlul Hoque, Ian Marquette and Yao-Zhong Zhang}

\address{School of Mathematics and Physics, The University of Queensland, Brisbane, QLD 4072, Australia}

\ead{m.hoque@uq.edu.au; i.marquette@uq.edu.au; yzz@maths.uq.edu.au}

\begin{abstract}
Classical and quantum superintegrable systems have a long history and they possess more integrals of motion than degrees of freedom. They have many attractive properties, wide applications in modern physics and connection to many domains in pure and applied mathematics. We overview two new families of superintegrable Kepler-Coulomb systems with non-central terms and superintegrable Hamiltonians with double singular oscillators of type $(n,N-n)$ in $N$-dimensional Euclidean space. We present their quadratic and polynomial algebras involving Casimir operators of $so(N+1)$ Lie algebras that exhibit very interesting decompositions $Q(3)\oplus so(N-1)$, $Q(3)\oplus so(n)\oplus so(N-n)$ and the cubic Casimir operators. The realization of these algebras in terms of deformed oscillator enables the determination of a finite dimensional unitary representation. We present algebraic derivations of the degenerate energy spectra of these systems and relate them with the physical spectra obtained from the separation of variables. 
\end{abstract}

\section{Introduction}
Algebraic methods are powerful tools in modern physics. Well known examples include the $N$-dimensional hydrogen atom and harmonic oscillator which were studied using the $so(N+1)$ \cite{Barr1,Lou1} and $su(N)$ \cite{Lou2,Hwa1} Lie algebras respectively. In particular the spectrum of the 5D hydrogen atom have been calculated using its $so(6)$ Lie algebra and the corresponding Casimir operators of order two, three and four \cite{Tru1}. Superintegrable models are an important class of quantum systems which can be solved using algebraic approaches. An important property of such systems is the existence of non-Abelian symmetry algebras generated by integrals of motion. These symmetry algebras can be embedded in certain non-invariance algebras involving non-commuting operators. Such symmetry algebras are in general finitely generated polynomial algebras and only exceptionally finite dimensional Lie algebras.

Quadratic algebras have been used to obtain energy spectrum of superintegrable systems \cite{Gra1}. The structure of a class of quadratic algebras with only three generators was studied and applied to 2D superintegrable systems in \cite{Das1}. Many researchers have studied quadratic and polynomial symmetry algebras of superintegrable systems and their representation theory ( see e.g. \cite{Kal1,Kal2, Das2, Win2,Isa1, Vin1, Vin2, Que1}).  

We review the results of our recent two papers \cite{FH1,FH2} and present the algebraic derivation of the complete energy spectrum of the $N$-dimensional superintegrable Kepler-Coulomb system with non-central terms and superintegrable double singular oscillators.

\section{The main results}
Let us consider the $N$-dimensional superintegrable Kepler-Coulomb system with non-central terms and superintegrable Hamiltonian with double singular oscillators of type $(n, N-n)$ introduced in \cite{FH1,FH2} 
\begin{eqnarray}
&&H_{KC}=\frac{1}{2}p^{2}-\frac{c_{0}}{r}+c_1\chi_1(r,x_N)+c_2\chi_2(r,x_N), \label{hamil1}
\\&&
H_{dso}=\frac{p^2}{2}+\frac{\omega^2 r^2}{2}+c_1\phi_1(x_1,\dots,x_n)+c_2\phi_2(x_{n+1},\dots,x_{N}),\label{hamil2}
\end{eqnarray}
where $ \vec{r}=(x_{1},x_{2},...,x_{N})$, $\vec{p}=(p_{1},p_{2},...,p_{N})$, $r^{2}=\sum_{i=1}^{N}x_{i}^{2}$, $p_{i}=-i \hbar \partial_{i}$, $\chi_1(r,x_N)=\frac{1}{r(r+x_N)}$, $\chi_2(r,x_N)=\frac{1}{r(r-x_N)}$, $\phi_1(x_1,\dots,x_n)=\frac{1}{x_1^2+\dots+x^2_n}$, $\phi_2(x_{n+1},\dots,x_{N})=\frac{1}{x^2_{n+1}+\dots+x^2_{N-n}}$ and $c_0$, $c_1$, $c_2$ are positive real constants. The system (\ref{hamil1}) is a generalization of the 3D superintegrable system in $E_3$ \cite{Kib1}. It includes as a particular 3D case the Hartmann potential which has applications in quantum chemistry and its classical analog possesses closed trajectories and periodic motion \cite{Tru1}. The model (\ref{hamil2}) is the generalization of the 4D and 8D systems obtained via the Hurwitz transformation for which only the symmetry cases (2,2) and (4,4) were studied \cite{Hur1, Mar1, Mar2}.

The model (\ref{hamil1}) is multiseparable and allows separation of variables in hyperspherical and hyperparabolic coordinates. The wave function is 
\begin{equation}
\psi(r,\Omega)=R(r)\Theta(\Phi_{N-1})y(\Omega_{N-2})
\end{equation}
in terms of special functions
\begin{eqnarray}
&&R(r)\propto(\varepsilon r)^{l+\frac{\delta_{1}+\delta_{2}}{2}} e^{\frac{-\varepsilon r}{2}}{}_1 F_1(-n+l+1, 2l+\delta_{1}+\delta_{2}+N-1; \varepsilon r),
\\&&
\Theta(z)\propto(1+z)^{\frac{(\delta_{1}+I_{N-2})}{2}}(1-z)^{\frac{(\delta_{2}+I_{N-2})}{2}} P^{(\delta_{2}+I_{N-2}, \delta_{1}+I_{N-2})}_{l-I_{N-2}}(z),
\\&&
\Lambda^2(N-1)y(\Omega_{N-2})=I_{N-2}(I_{N-2}+N-3)y(\Omega_{N-2}),\label{fk3}
\end{eqnarray}
where $z=\cos(\Phi_{N-1})$, $I_{N-2}(I_{N-2}+N-3)$ being the general form of the separation constant,
$P^{(\alpha, \beta)}_{\lambda}$ denotes a Jacobi polynomial and $l\in{\rm I\!N}$, $\delta_{i}=\{\sqrt{(I_{N-2}+\frac{N-3}{2})^2+4c'_{i}}-\frac{N-3}{2}\}-I_{N-2},c'_i=\frac{c_i}{\hbar^2}, i=1, 2$. The energy spectrum of the system (\ref{hamil1}) is \cite{FH1}
\begin{eqnarray}
E_{KC}\equiv E_n=\frac{-c^{2}_{0}}{2\hbar^{2}\left( n+\frac{\delta_{1}+\delta_{2}}{2}+\frac{ N-3}{2}\right)^{2}},\quad n=1,2,3,\dots\label{en2}
\end{eqnarray}
The system (\ref{hamil2}) is also multiseparable and allows separation of variables in double hyper Eulerian and double hyperspherical coordinates. We can split (\ref{hamil2}) into the sum of two singular oscillators of dimensions $n$ and $N-n$ as $H_{dso}=H_1+H_2$, where 
\begin{eqnarray}
 &&H_1=\frac{1}{2}(p^2_1+...+p^2_n)+\frac{\omega^2 r^2_1}{2}+c_1\phi_1(x_1,\dots,x_n),
 \\&&
 H_2=\frac{1}{2}(p^2_{n+1}+...+p^2_N)+\frac{\omega^2 r^2_2}{2}+c_2\phi_2(x_{n+1},\dots,x_{N}).
 \end{eqnarray}
 The wave function $\psi(r_1,\Omega)=R_1(r_1)y(\Omega_{n-1})$ of $H_1$ in terms of special functions is given by
 \begin{eqnarray}
\psi_{n_1 l_n}(u)\propto\frac{a e^{-\frac{u}{2}} u^{(\delta_1+\frac{1}{2}l_{n})/2}}{\sqrt{2(\delta_1+\frac{1}{2}l_{n}+\frac{n}{4})}} \times{}_1F_1\left(-n_1; 2\{\delta_1+\frac{1}{2}l_{n}+\frac{n}{4}\}; u\right),\label{kpWf}
\end{eqnarray}
where 
$\delta_1=\left\{\sqrt{(\frac{1}{2}l_{n}+\frac{n-2}{4})^2+\frac{1}{2}c'_1}-\frac{n-2}{4}\right\}-\frac{1}{2}l_{n}$, $n_1=\frac{E'_1}{2\omega'}-\left(\delta_1+\frac{1}{2}l_{n}+\frac{n}{4}\right)$, $c'_1=\frac{c_1}{\hbar^2}$, $\omega'=\frac{\omega}{\hbar}$ and $E'_1=\frac{E_1}{\hbar^2}$. The wave equation of $H_2$ in $(N-n)$-dimensional hyperspherical coordinates has similar solution. The energy spectrum of the system (\ref{hamil2}) is \cite{FH2} 
\begin{eqnarray}
E_{dso}=2\hbar \omega\left(p+1+\frac{\alpha_1+\alpha_2}{2}\right)\label{Edso},
\end{eqnarray}
where $\alpha_1=2\delta_1+l_{n}+\frac{n-2}{2}$, $\alpha_2=2\delta_2+l_{N-n}+\frac{N-n-2}{2}$ and $p=n_1+n_2$.

The integrals of motion of (\ref{hamil1}) are given by
\begin{eqnarray}
&A&=\sum_{i<j}^{N}L_{ij}^{2}+ 2r^2[c_1\chi_1(r,x_N)+c_2\chi_2(r,x_N)],\label{kf5}
\\&B&=-M_{N}+c_{1}(r-x_{N})\chi_1(r,x_N)-c_{2}(r+x_{N})\chi_2(r,x_N),\label{kf6}
\\&J^{2}&=\sum_{i<j}^{N-1}L_{ij}^{2},\quad L_{i j}=x_{i}p_{j}-x_{j}p_{i}, \quad  i, j=1, ... , N-1,
\end{eqnarray}
and for the model (\ref{hamil2}) they are
\begin{eqnarray}
&A&=-\frac{h^2}{4}\left\{\sum^N_{i,j=1}x^2_i\partial^2_{x_j}-\sum^N_{i,j=1}x_i x_j \partial_{x_i}\partial_{x_j}-(N-1)\sum^N_{i=1}x_i \partial_{x_i}\right\}\nonumber\\&& +\frac{c_1\phi_1(x_1,\dots,x_n)+c_2\phi_2(x_{n+1},\dots,x_{N})}{2}\left[\frac{1}{\phi_1(x_1,\dots,x_n)}+\frac{1}{\phi_2(x_{n+1},\dots,x_{N})}\right],\label{kpA}
\\
&B&=\frac{1}{2}\left\{\sum^n_{i=1}p^2_i-\sum^N_{i=n+1}p^2_i\right\} + \frac{\omega^2}{2}\left\{\frac{1}{\phi_1(x_1,\dots,x_n)}-\frac{1}{\phi_2(x_{n+1},\dots,x_{N})}\right\}\nonumber\\&&\quad+c_1\phi_1(x_1,\dots,x_n)-c_2\phi_2(x_{n+1},\dots,x_{N}),\label{kpB}
\\
&J_{(2)}&=\sum_{i<j}J^2_{ij},\quad J_{ij}=x_i p_j-x_j p_i, i, j= 1,2,....,n, 
\\
&K_{(2)}&=\sum_{i<j}K^2_{ij},\quad K_{ij}=x_i p_j-x_j p_i, i, j= n+1,....,N,\label{kpJ}
\end{eqnarray}
where the Runge-Lenz vector
$M_{j}=\frac{1}{2} \sum_{i=1}^{N} ( L_{ji}p_{i}-p_{i}L_{ij} )- \frac{c_{0}x_{j}}{r}$.
The integrals of motion \{$A,B,C$\} close to the general form of the quadratic algebra $Q(3)$ \cite{Das1},
\begin{eqnarray}
&[A,B]&=C,\label{prova5}
\\&[A,C]&=\alpha A^2+\gamma \{A,B\}+\delta A+\epsilon B+\zeta,\label{prova6}
\\&[B,C]&=aA^2-\gamma B^{2}-\alpha\{A,B\} +d A -\delta B + z\label{prova1}
\end{eqnarray}
and the Casimir operator
\begin{eqnarray}
&K&=C^{2}-\alpha\{A^2,B\}-\gamma \{A,B^{2}\}+(\alpha \gamma-\delta)\{A,B\}+ (\gamma^2-\epsilon) B^{2} \nonumber\\&&+ (\gamma\delta-2\zeta) B+\frac{2a}{3}A^3 +(d+\frac{a\gamma}{3}+\alpha^2) A^{2} +(\frac{a\epsilon}{3}+\alpha\delta+2z)A.\label{prova4} 
\end{eqnarray}
Here coefficients $\alpha, \gamma,\delta, \epsilon, \zeta, a, d, z$ are given in the following table for models $H_{KC}$ and $H_{dso}$ \cite{FH1,FH2}.
\begin{table}[h] 
\begin{center}
    \begin{tabular}{ | l | p{6cm} | p{6cm} |}
    \hline
    &${\bf H_{KC}}$ & ${\bf H_{dso}}$ 
    \\ \hline
     $\alpha$ & 0 & 0 
    \\ \hline
   $\gamma$ & $2 \hbar^{2}$ & $2 \hbar^{2}$ 
    \\ \hline
     $\delta$ & 0 &  0
     \\ \hline
     $\epsilon$ &$(N-1)(N-3) \hbar^{4}$& $\frac{\hbar^4}{4}N(N-4)$
         \\ \hline
     $\zeta$ & $-4 (c_{1}-c_{2}) \hbar^{2} c_{0}$ &  $-\hbar^2 J_{(2)}H+\hbar^2 K_{(2)}H -\frac{\hbar^2}{4}\{8c_1-8c_2-(N-4)(N-2n)\hbar^2\}H$
         \\ \hline
     $a$ & 0 &  0
         \\ \hline
     $d$ & $8 \hbar^{2} H$ & $-16\hbar^2\omega^2$
         \\ \hline
     $z$ & $-4 \hbar^{2} J^{2} H + (N-1)^{2} \hbar^{4} H-8 \hbar^{2}(c_{1}+c_{2})H + 2 \hbar^{2} c_{0}^{2} $ &  $2\hbar^2 H^2+4\hbar^2\omega^2 J_{(2)}+4\hbar^2\omega^2 K_{(2)}+8\hbar^2\omega^2\{c_1+c_2-\frac{\hbar^2}{4}n(N-n)\}$
    \\
    \hline
    \end{tabular}
    \caption{\label{tabone} Coefficients in $Q(3)$ and $K$.}
\end{center}
\end{table}
\\
By means of the explicit expressions of $A, B, C$, we can write the Casimir operator in terms only of central elements as
\begin{eqnarray}
&K_{KC}&=2(N-3)(N-1)\hbar^{4} H J^{2} -8 \hbar^{2}(c_{1}-c_{2})^{2} H + 4  (N-3) (N-1) (c_{1}+c_{2}) \hbar^{4} H 
 \nonumber\\&&\quad-\hbar^{6}(N-3)(N-1)^{2} H +4 \hbar^{2} c_{0}^{2} J^{2} +8 \hbar^{2} (c_{1}+c_{2})c_{0}^{2} -2 (N-3) \hbar^{4} c_{0}^{2},\label{prova3} 
\end{eqnarray}
\begin{eqnarray}
&K_{dso}&=2\hbar^2 J_{(2)}H^2+2\hbar^2 K_{(2)}H^2+\frac{\hbar^2}{4}\left[16c_1+16c_2-\{4(N-4)-(N-2n)^2\}\hbar^2\right]H^2\nonumber\\&&+\hbar^2\omega^2 J^2_{(2)}+\hbar^2\omega^2 K^2_{(2)}-2\hbar^2\omega^2 J_{(2)} K_{(2)}+4\hbar^2\omega^2\{c_1-c_2-\frac{1}{4}(N-4)(N-n)\hbar^2\}J_{(2)}\nonumber\\&&-4\hbar^2\omega^2\{c_1-c_2+\frac{1}{4}n(N-4)\hbar^2\}K_{(2)}+4\hbar^2\omega^2\left[(c_1-c_2)^2-\frac{1}{2}(N-n)(N-4)\hbar^2 c_1\right.\nonumber\\&&\left.-\frac{1}{2}n(N-4)\hbar^2 c_2+\frac{1}{4}n(N-n)(N-4)\hbar^4\right].\label{kpK1}
\end{eqnarray}
The first order integrals $L_{ij}$, $i, j, k, l=1, .., N-1$ of $H_{KC}$ generate $so(N-1)$ Lie algebra 
\begin{eqnarray}
[L_{ij},L_{kl}]=i ( \delta_{ik}L_{jl}+ \delta_{jl}L_{ik}-\delta_{il}L_{jk}-\delta_{jk}L_{il})\hbar
\end{eqnarray}
and
$J_{ij}=x_i p_j-x_j p_i$, $i, j= 1,2,....,n$ and 
$K_{ij}=x_i p_j-x_j p_i$, $i, j= n+1,....,N$ of $H_{dso}$ generate $so(n)$ and $so(N-n)$ Lie algebras
\begin{eqnarray}
&[J_{ij},J_{kl}]&= i(\delta_{ik}J_{jl}+ \delta_{jl}J_{ik}-\delta_{il}J_{jk}-\delta_{jk}J_{il})\hbar, 
\\&[K_{ij},K_{kl}]&=i( \delta_{ik}K_{jl}+ \delta_{jl}K_{ik}-\delta_{il}K_{jk}-\delta_{jk}K_{il})\hbar.
\end{eqnarray}
Thus the full symmetry algebra is $Q(3)\oplus so(N-1)$ for $H_{KC}$ and $Q(3)\oplus so(n)\oplus so(N-n)$ for $H_{dso}$. Chains of second order Casimir operators associated with $so(N-1)$, $so(n)$ and $so(N-n)$ components may be used to label quantum states for  $H_{KC}$ and $H_{dso}$ respectively. 

These quadratic algebras can be realized in terms of the deformed oscillator algebra \cite{Das1,Das3} \begin{eqnarray}
[\aleph,b^{\dagger}]=b^{\dagger},\quad [\aleph,b]=-b,\quad bb^{\dagger}=\Phi (\aleph+1),\quad b^{\dagger} b=\Phi(\aleph),
\end{eqnarray}
where $\aleph$ is the number operator and $\Phi(x)$ is a well behaved real function satisfying $\Phi(x)>0$ for all $x>0$. Then the structure function has the form \cite{FH1,FH2} 
\begin{eqnarray}
\Phi(x; u,E)=\nu_0\prod^6_{i=1} [x+u-\nu_i],\label{St}
\end{eqnarray}
where $\nu_i$ are given in the following table.
\\
\begin{table}[h] 
\begin{center}
    \begin{tabular}{ | l | l | l |}
    \hline
    &${\bf H_{KC}}$ & ${\bf H_{dso}}$ 
    \\ \hline
     $\nu_0$ & 6291456$E\hbar^{18}$ & -12582912$\hbar^{18}\omega^2$
    \\ \hline
   $\nu_1$ & $\frac{1}{2}(1+m_{1}+m_{2})$ & $\frac{1}{4}(2+m'_{1}+m'_{2})$ 
    \\ \hline
     $\nu_2$ & $\frac{1}{2}(1+m_{1}-m_{2})$ &  $\frac{1}{4}(2+m'_{1}-m'_{2})$
     \\ \hline
     $\nu_3$ &$\frac{1}{2}(1-m_{1}+m_{2})$& $\frac{1}{4}(2-m'_{1}+m'_{2})$
         \\ \hline
     $\nu_4$ &$\frac{1}{2}(1-m_{1}-m_{2})$ &  $\frac{1}{4}(2-m'_{1}-m'_{2})$
         \\ \hline
     $\nu_5$ & $\frac{1}{2}+\frac{c_0}{\hbar\sqrt{-2E}}$ &  $\frac{\hbar\omega+H}{2\hbar\omega}$
         \\ \hline
     $\nu_6$ & $\frac{1}{2}-\frac{c_0}{\hbar\sqrt{-2E}}$ & $\frac{\hbar\omega-H}{2\hbar\omega}$
            \\
    \hline
    \end{tabular}
 \caption{\label{tabone} Coefficients $\nu_i$ in (\ref{St}). }
\end{center}
\end{table}
\\
In the above table, $\hbar^2 m_{1,2}^{2} =16 c_{1,2}+\{4J^2 +(N-3)^{2}\}\hbar^2$, $\hbar^2 m'^2_1 =8 c_1+4J_{(2)} +(n-2)^2\hbar^2$ and $\hbar^2 m'^2_2 =8 c_2+4K_{(2)} +(N-n-2)^2\hbar^2$.

For a finite-dimensional unitary representation, we have the following constraints
\begin{equation}
\Phi(p+1; u, E)=0;\quad \Phi(0;u,E)=0;\quad \Phi(x)>0,\quad 0<x<p+1\label{pro2}
\end{equation}
on the structure function (\ref{St}). The solutions of these constraints provide the energy spectra $E_{KC,dso}$ and the arbitrary constant $u_{KC, dso}$ ($\epsilon_1=\pm 1, \epsilon_2=\pm 1$) as 
\begin{eqnarray}
&&E_{KC}=\frac{-2 c^2_{0}}{h^2 (2 + 2 p +\epsilon_{1} m_{1} +\epsilon_{2} m_{2} )^2},\quad u_{KC}=\frac{1}{2}+\frac{c_{0}}{\hbar\sqrt{-2E}},
\\
Case\quad 1:&& E_{dso}=2\hbar\omega(p+1+\frac{\epsilon_{1} m'_{1} +\epsilon_{2} m'_{2}}{4}),\qquad u_{dso}=\frac{ E+\hbar\omega}{2\hbar\omega},
\\
Case\quad 2:&& E_{dso}=2\hbar\omega(p+1+\frac{\epsilon_{1} m'_{1} +\epsilon_{2} m'_{2}}{4}),\qquad u_{dso}=\frac{- E+\hbar\omega}{2\hbar\omega},
\\
Case\quad 3:&& E_{dso}=2\hbar\omega(p+1+\frac{\epsilon_{1} m'_{1} +\epsilon_{2} m'_{2}}{4}),\qquad u_{dso}=\frac{1}{4}(2+\epsilon_1 m_1+\epsilon_2 m_2).
\end{eqnarray}
We have four possible structure functions for $H_{KC}$ \cite{FH1} and twenty four for $H_{dso}$ \cite{FH2}. 
Making the identification $n_1+n_2+I_{N-2}=n-1$, $p=n_{1}+n_{2}$, $m_{i}=\frac{1}{2}(3-2I_{N-2}-N-2\delta_{i}), m'_i=2\alpha_i, i=1, 2$, the energy spectra $E_{KC}$ and $E_{dso}$ coincide (\ref{en2}) and (\ref{Edso}) respectively. 

\section{Conclusion}
We have presented some of the results in \cite{FH1,FH2} and shown how the $su(N)$ and $so(N+1)$ symmetry algebras of the $N$-dimensional Kepler-Coulomb (i.e., $c_1=c_2=0$) and the $N$-dimensional harmonic oscillator (i.e., $c_1=c_2=0$) are broken to higher rank polynomial algebra of the form $Q(3)\oplus L_1\oplus L_2\oplus\dots$ for non-zero $c_1$ and $c_2$, where $L_1, L_2,\dots$, are certain Lie algebras. $Q(3)$ is a quadratic algebra involving Casimir operators of the Lie algebras in its structure constants.  We have also presented the realizations of these quadratic algebras in terms of deformed oscillator algebras and obtained the finite dimensional unitary representations which yield the energy spectra of these superintegrable systems. In addition, we have compared them with the physical spectra obtained from the separation of variables.

The systems (\ref{hamil1}) and (\ref{hamil2}) could be studied using new approaches such as the recurrence method related to special functions and orthogonal polynomials and approaches combining ladder, shift, intertwining and supercharges operators \cite{Mar1,Mill2}. The generalizations of these systems to include monopole interactions and their duals \cite{Mar2,Mard5} could be also investigated.
 
{\bf Acknowledgements:}
The research of FH was supported by International Postgraduate Research Scholarship and Australian Postgraduate Award. IM was supported by the Australian Research Council through a Discovery Early Career Researcher Award DE 130101067. YZZ was partially supported by the Australian Research Council, Discovery Project DP 140101492.

\end{document}